\documentclass[a4,11pt]{article}

\usepackage{amsmath,amssymb,color,graphics,amscd,amsfonts,epsf}
\allowdisplaybreaks[4]

\date{}
\begin{document}

\begin{flushright} 
 
TAUP-2933/11
\end{flushright} 

\vspace{0.1cm}

\begin{center}
  {\LARGE 
On a new type of orbifold equivalence and M-theoretic $AdS_4/CFT_3$ duality

  }
\end{center}
\vspace{0.1cm}
\vspace{0.1cm}
\begin{center}

         Masanori H{\sc anada}$^{a}$\footnote
         {
E-mail address : mhanada@u.washington.edu}, 
 Carlos H{\sc oyos}$^{a,b}$\footnote
         {
E-mail address : choyos@phys.washington.edu}, 
and 
Hidehiko S{\sc himada}$^{c}$\footnote
         {
E-mail address : shimada@nbi.dk}

\vspace{0.3cm}

${}^a$ {\it Department of Physics, University of Washington, 

 Seattle, WA 98195-1560, USA}\\

 ${}^{b}$ 
{\it Raymond and Beverly Sackler School of Physics and Astronomy

Tel-Aviv University, Ramat-Aviv 69978, Israel}\\

${}^{c}$ 
{\it  Niels Bohr International Academy, The Niels Bohr Institute, University of Copenhagen,

 Blegdamsvej 17, DK-2100 Copenhagen, Denmark}

\end{center}

\vspace{1.5cm}

\begin{center}
  {\bf abstract}
\end{center} 

We consider the large-$N$ limit of 
$\mathcal{N}=6$ $U(N) \times U(N)$ 
superconformal Chern-Simons (ABJM) theory
with fixed level $k$, 
which is conjectured to be dual to M-theory on $AdS_4\times (S^7/{\mathbb Z}_k)$ background. 
We point out that the so-called orbifold equivalence on the gravity side, 
combined with the $AdS_4/CFT_3$ duality, predicts
a hitherto unknown type of duality on the gauge theory side.
It establishes the equivalence between a class of observables,
which are not necessarily protected by supersymmetry,
in strongly coupled ABJM theories away from the planar approximation,
with different values of $k$ and $N$ but sharing common $kN$.  
This limit is vastly different from the planar limit, 
and hence from the gauge theory point of view 
the duality is more difficult to explain 
compared to the previously known analogous equivalence between planar gauge theories, 
where one can explicitly prove the equivalence diagrammatically using 
the dominance of the planar diagrams. 
 
 \newpage

\section*{Introduction and main result}

Three-dimensional gauge theories with a Chern-Simons term have been a subject of interest 
in theoretical physics ever since their first appearance~\cite{Deser:1982vy, Deser:1981wh}. 
They have direct applications in condensed matter physics
and exhibit rich mathematical structure.
In recent years Chern-Simons theories coupled to matter fields,
with various amount of supersymmetry, have gained renewed interest because they 
capture some part of the dynamics of the elusive M-theory, the eleven-dimensional
theory defined by the strong coupling limit of type IIA superstring theory.  
In particular Aharony, Bergman, Jafferis and Maldacena constructed an  
$\mathcal{N}=6$ $U(N) \times U(N)$ superconformal Chern-Simons theory 
(ABJM theory)
with a certain coupling constant parametrised by an integer (the level) $k$~\cite{Aharony:2008ug},  
following important earlier works~\cite{Schwarz:2004yj, Bagger:2007jr}.  
The model 
has been proposed to be the low energy effective theory of $N$ M2 branes (which are important
degrees of freedom in M-theory) put on an orbifolded (transverse) space $\mathbb{R}^8/{\mathbb Z}_k$.
It has been conjectured that the low energy effective theory of $N$ M2 branes in flat space $\mathbb{R}^8$ should be dual to M-theory in $AdS_4 \times S^7$ when $N$ is large~\cite{Maldacena:1997re}, 
and in~\cite{Aharony:2008ug}, accordingly, it was conjectured that the ABJM theory
is dual to M-theory on $AdS_4 \times S^7/{\mathbb Z}_k$.  

Similar orbifold geometries have been 
considered in the context of the $AdS_5/CFT_4$ duality in 
\cite{Kachru:1998ys}, where 
the duality between type IIB string theory in $AdS_5 \times S^5/{\mathbb Z}_k$
and the corresponding Yang-Mills theories in four dimensions was discussed.
It was found that, from the gravity side, one can predict 
an equivalence between these field theories for observables
which are invariant under the ${\mathbb Z}_k$ projection 
(up to a calculable factor depending on $k$). 
The equivalence holds only at tree level on the gravity side.
This is because for tree level processes, all intermediate states are also 
invariant under ${\mathbb Z}_k$, whereas ${\mathbb Z}_k$ breaking intermediate   
states can appear in loop corrections. 

In this note we will discuss the application of 
this idea to the $AdS_4/CFT_3$ duality. 
We will show that similar arguments lead to predictions for
the dual gauge theory which are more surprising and 
difficult to explain using field theoretical methods. 
Let us consider two ABJM theories with levels $k_1, k_2$ and  
gauge groups $U(N_1) \times U(N_1)$, $U(N_2)\times U(N_2)$. 
The curvature radius of the dual background in eleven-dimensional Planck units 
is given by~\cite{Aharony:2008ug}
\begin{equation}
\frac{R}{l_p}
=(2^5 \pi^2 N^\prime_i)^{\frac{1}{6}}, 
\label{11d_curvature}
\end{equation}
where $N^\prime_i\equiv N_ik_i$.~\footnote{In addition to the metric 
there is also four-form flux, but it depends on the parameters of the theory only through $R$,
so the full supergravity solution depends only on $N^\prime$.
}
Hence, for theories with the same value of $N^\prime$,
\begin{equation}
N_1 k_1 = N^\prime = N_2 k_2, 
\end{equation} 
any observables which are blind both to 
$\mathbb{Z}_{k_1}$ and $\mathbb{Z}_{k_2}$ transformations
should be equal in the two theories, 
as long as loop corrections on the gravity side are negligible.~\footnote{
We notice that in general 
there are observables, 
say in the first theory dual to the $\mathbb{Z}_{k_1}$ orbifold, 
that are charged under the $\mathbb{Z}_{k_2}$ transformation 
associated to the second orbifold. They are absent in the second ABJM theory, and
hence there are no orbifold equivalence for these operators.}
There are many observables which are blind to the $\mathbb{Z}_{k_1}$ and $\mathbb{Z}_{k_2}$ transformations,
both supersymmetric and non-supersymmetric.~\footnote{For BPS operators, 
the orbifold equivalence
may be considered as a simple consequence of the usual prescrition to compute correlation functions in $AdS/CFT$~\cite{Gubser:1998bc, Witten:1998qj}.
In the $AdS_5/CFT_4$ case, an extension to near-BPS operators is
discussed in~\cite{Shimada:2004sw}. Orbifold equivalence for non-BPS operators might be
understood along this line.}
We discuss some examples later.

Let us discuss the conditions under which quantum corrections are negligible. 
In general this depends strongly on the observable (or the process) 
we consider. 
However, one can give necessary conditions: 
if length scales involved in the model (on the gravity side)
are comparable to the eleven-dimensional Planck scale, 
one cannot expect quantum gravity corrections to be small. 
There are two length scales in the geometry, 
the curvature radius of the background $R$ given in \eqref{11d_curvature} 
and the radius of the M-theory circle $R/k$. 
Requiring both of them to be much larger than the Planck scale, one obtains the conditions 
$N'=Nk\gg 1$ and $N\gg k^5$, where the latter 
implies the former (because $k, N$ are integers)~\cite{Aharony:2008ug}.

The argument given so far,
which is on the gravity side, is completely parallel to that in the $AdS_5/CFT_4$ duality. 
However, from the gauge theory point of view we find a big difference.
In the context of $AdS_5/CFT_4$ duality, 
it is possible to prove the equivalence 
directly in the gauge theory~\cite{Bershadsky:1998cb}; 
the essential point in the proof is that 
loop effects on the gravity side correspond to $1/N$ corrections on the gauge theory side. 
Hence one should focus on planar diagrams, and one can indeed prove
the equivalence using purely field theoretic arguments.  
However, in our case the same type of argument is not applicable. 
The reason is as follows. 
In ABJM theories, the 't Hooft expansion is the double expansion in terms of $1/N$ and the 't Hooft coupling $\lambda=N/k$. 
The dominance of planar diagrams
holds in the large-$N$ limit with fixed $\lambda$, which is dual to type IIA superstring in $AdS_5\times {\mathbb CP }^3$~\cite{Aharony:2008ug}. 
However, in the M-theory regime, $k$ is typically of order one and  $N$ is still large, so that  
neglecting contributions from non-planar diagrams
cannot be justified.
Thus our conclusion is that the $AdS_4/CFT_3$ duality, combined with the orbifold equivalence,
predicts non-trivial relations for very strongly coupled 
(in the sense that the planar approximation cannot be used) ABJM theories. 
This new equivalence is much more difficult to explain in field theoretic terms
than the analogous equivalence relations 
found in the $AdS_5/CFT_4$ context (and their extensions such as those considered in
\cite{Cherman:2010jj}).~\footnote{
Of course, one can consider the more familiar type of 
the orbifold equivalence associated with planar diagrams in the 
$AdS_4/CFT_3$ context as well, but in the limit where $N/k$ is fixed,
not in the M-theoretic limit we are interested in. 
For these planar orbifold equivalence, 
the internal space ${\mathbb CP}^3$ is orbifolded on the gravity side. 
On the gauge theory side, the gauge group changes 
as in the $AdS_5/CFT_4$ case; see e.g.~\cite{Benna:2008zy}.  }
\section*{Observables} 

Because in the region of parameters that we are considering
the gauge theory is strongly coupled, 
it is not easy to explicitly compute observables.
However, recent developments using the localization technique  
provide us with some exact results
in ABJM theories.
In particular, the ``free energy'' $\log Z$ of the theory in a compact three-sphere $S^3$
in the planar limit ($N, k \to \infty$ with $N/k$ fixed),
and some sub-leading terms in the $1/N$ expansion were 
obtained in ~\cite{Drukker:2010nc, Drukker:2011zy}.
Let us discuss the free energy using 
an expression 
proposed in~\cite{Fuji:2011km} 
which sums up all $1/N$ corrections 
\begin{equation}
\log Z = \log \left( 
2\pi C_1 Ai\left(\left(\frac{\pi}{\sqrt{2}} \left(\frac{N}{\lambda}\right)^2
\lambda_{ren}^{3/2}\right)^{2/3} \right)\right), 
\label{RFlogZ}
\end{equation}
where $Ai(x)$ is the Airy function, 
\begin{equation}
C_1=\frac{1}{\sqrt{2}}\left(\frac{2\pi}{k}\right)^{-1/3}
\end{equation}
and the `renormalized 't Hooft coupling' $\lambda_{ren}$ is given by  
\begin{equation}
\lambda_{ren}=\lambda-\frac{1}{24} -\frac{\lambda^2}{3N^2}. 
\end{equation}
Whether this formula is applicable in the M-theory regime is not very clear\footnote{
\textbf{Note added:} 
After the first version of this paper appeared in the arXiv,
numerical results are obtained which support the validity of
\eqref{RFlogZ} in the M-theory region \cite{HHHNSY}.  See also \cite{arXiv:1110.4066}.  
},
as the limit $N\to \infty$ with $k$ fixed is very different from the planar limit.
However, a direct extrapolation of this expression to the M-theory region
provides us with a result which is consistent with the orbifold equivalence. 
The asymptotic behavior of the Airy function at a large positive value of $x$
\begin{equation}
Ai(x)\sim \frac{1}{2\sqrt{\pi}}\left(\frac{1}{x}\right)^{1/4}\exp{\left(-\frac{2}{3} x^{\frac{3}{2}}\right)}
\end{equation} 
yields, to leading order,
\begin{equation}
\log Z
\simeq
-\frac{\sqrt{2}\pi}{3} 
k^{1/2}N^{3/2}
=
-\frac{\sqrt{2}\pi}{3}
\frac{N^{\prime 3/2}}{k}. \label{RFLEADINGFHM}
\end{equation} 
The $1/k$ scaling of the last expression is what is expected from the orbifold equivalence 
-- the free energy is proportional to  
the volume of $S^7/{\mathbb Z}_k$ on the gravity side.  
The leading order term in the asymptotic form of the Airy function 
in the M-theory regime ($k$ finite, $N\to \infty$) is
actually the same as the type IIA supergravity limit, 
where $\lambda \to \infty$ is taken after the $N \to \infty$
limit with fixed $\lambda=\frac{N}{k}$. 
This is an expected result from the gravity side~\cite{Aharony:2008ug} and
is first derived on the field theory side in~\cite{Drukker:2011zy}
by rewriting the 't Hooft expansion in terms of M-theory variables. 
One may also read off the sub-leading correction 
in this regime 
(which is first obtained in ~\cite{Drukker:2011zy}) 
from (\ref{RFlogZ}),~\footnote{
In~\cite{Fuji:2011km}, a discrepancy between the sub-leading correction in this formula 
and a calculation in string theory~\cite{Bergman:2009zh} was pointed out.
These terms do not affect the leading order 
expression \eqref{RFLEADINGFHM}. 
} 
\begin{eqnarray}
\log Z 
&=&
-\frac{\sqrt{2}\pi}{3}
\frac{N^{\prime 3/2}}{k}
\left(
1-\frac{1}{2N^\prime}
-\frac{k}{16N^\prime}
\right). 
\end{eqnarray}
We note that the last term which does not satisfy
the expected $1/k$ behaviour should be interpreted as coming from quantum correlations in $S^7/{\mathbb Z}_k$.

Another class of observables comprises 
an infinite family of local BPS operators 
\cite{Aharony:2008ug}
\begin{eqnarray}
Tr\left(
C_{I_1}C^\dagger_{J_1}\cdots C_{I_l}C^\dagger_{J_l}
\right), \label{RFCoperator}
\end{eqnarray}
where $I_1,\cdots,I_l$ and $J_1,\cdots,J_l$ are symmetrized.
Here, $C_I\ (I=1,2,3,4)$ denote four complex bi-fundamental 
scalar fields in ABJM theories
which describe the collective coordinates
of M2-branes~\footnote{
Our notation is that of the original paper~\cite{Aharony:2008ug}.}
, on which  the ${\mathbb Z}_k$ symmetry acts  as $C_I\to e^{2\pi i/k}C_I$.  
Operators \eqref{RFCoperator} are manifestly ${\mathbb Z}_k$ invariant. 
That their scaling dimensions are protected is of course consistent 
with the orbifold equivalence. 
There are other classes of gauge invariant 
BPS operators written in terms of monopole operators~\cite{Aharony:2008ug}.
One can choose the monopole charges $m_1$ and $m_2$ for the two theories
such that $k_1 m_1=k_2 m_2$ holds. These operators are 
invariant under both ${\mathbb Z}_{k_1}$ and ${\mathbb Z}_{k_2}$ symmetries. 
Our argument predicts that 
the equivalence should also hold for correlation functions of these operators
\footnote{We thank O.~Aharony for very useful comments on this issue.},
at tree level on the gravity side.
 
The observables we have discussed so far involve operators which
preserve some supersymmetry.
There are also non-supersymmetric, or non-BPS, observables 
which are ${\mathbb Z}_k$ invariant.~\footnote{
Non-BPS operators are not protected by supersymmetry and
hence equivalence between them would be more difficult to show
in the gauge theory. We note however that,
in ABJM theories (as opposed to the four-dimensional $\mathcal{N}=4$ Yang-Mills theory),
overall coefficients of even three-point correlation functions of BPS operators
are believed not to be protected, though
scaling dimensions are believed to be protected; see e.g.~\cite{Bhattacharya:2008bja}.}
The argument on the gravity side predicts equivalence of 
non-supersymmetric observables as well, 
as far as the tree level approximation 
on the gravity (M-theory) side
is justified.~\footnote{
It may be pointed out that we are implicitly assuming that M-theory
obeys usual principles of quantum mechanics; 
in particular we are assuming 
the existence of a semi-classical expansion, or the distinction 
between tree-level and loop-level processes in M-theory.}. 
One class of examples is provided by various observables at finite temperature. 
In this case, the gravity counterpart 
is calculated via a non-BPS solution to eleven-dimensional supergravity. 
Another example is the loop operator discussed in~\cite{Drukker:2008jm},
which is analogous to the Wilson-loop in the $AdS_5/CFT_4$ 
correspondence~\cite{Maldacena:1998im, Rey:1998ik}
The M-theory counterpart is the volume of the membrane minimal hyper-surface, 
with appropriate boundary conditions. 
For generic boundary conditions, 
these loop operators  
are non-BPS (although locally BPS).~\footnote{
We thank D. Young for a very useful
discussion on this point.}
In these cases, it is clear that
as far as the length scale involved in the solutions
 (of the supergravity equation in the finite-temperature physics, 
 and of the minimal hyper-surface equation for the loop operator)
is much larger than the Planck scale, 
the tree level approximation in M-theory should apply. 
In general, the dictionary of the AdS/CFT correspondence for 
non-BPS operators is not well established, compared to BPS operators.
Hence for some non-BPS observables, the validity of the tree level approximation is 
not immediately clear.

\section*{Conclusion and Discussion}

The most crucial point in our observation is that a very simple equivalence
on the gravity (M-theory) side translates into a highly non-trivial equivalence
on the gauge theory side.
This is because we have considered the M-theory regime where $k$ is kept finite
while $N$ is taken to be large. It is remarkable that one can predict
many relations in this limit, which is considered to be less under control 
than the 't Hooft limit, or equivalently the type IIA regime, where $N/k$ is fixed and 
both $k$ and $N$ are large.

Our argument brings in the following crucial question,
``how should one distinguish, 
on the gauge theory side, the tree and loop processes
in the dual gravity theory?''
or  
``what is the loop expansion parameter in the gravity theory, when
expressed in the language of the gauge theory?''
In the standard $AdS_5/CFT_4$ duality and the $AdS_4/CFT_3$ in the type IIA regime,
the answers are of course provided by the 't Hooft expansion~\cite{'tHooft:1973jz}: 
loop level processes (on the gravity side)
correspond to non-planar diagrams and the expansion parameter is $1/N$.
It is natural that these answers fail in the M-theory regime, as the 't Hooft expansion
implies that the dual theory is a theory of strings, whereas the fundamental 
degrees of freedom of eleven-dimensional M-theory are not strings.
It is clearly important to answer 
these questions in the M-theory regime; the orbifold
equivalence may well be useful here
as it provides a way
to discriminate the tree level and the loop level effects in 
M-theory. 

It is also important to find a way to understand the
equivalence in purely field theoretic terms, 
which is available for the usual planar orbifold equivalence~\cite{Bershadsky:1998cb}. 
Once one understands the equivalence without using the gravitational
dual, it may be possible to generate many examples of orbifold equivalence
between very strongly coupled Chern-Simons theories
including inherently non-supersymmetric ones. These equivalences,
in particular those between non-supersymmetric theories, 
may have applications in condensed matter.
Although supersymmetry is often considered to be crucial for the 
gauge/gravity duality, purely field theoretical orbifold equivalence
(in the planar limit)
holds without relying on supersymmetry.
Indeed, this idea was pursued in~\cite{Cherman:2010jj}
as a viable strategy to study finite density QCD theories,
circumventing the infamous sign problem. 

The orbifold equivalence considered in this paper
would provide a nontrivial test of the $AdS_4/CFT_3$ duality conjecture 
in M-theory,  
although checking it directly is not an easy problem.
The formulation of M-theory is not yet established,
and the dynamics of the best candidate for such a formulation
in flat spacetime, the matrix model~\cite{de Wit:1988ig, Banks:1996vh}, 
is far from being understood. M-theory dynamics on curved spacetime,
including $AdS_4 \times S^7$, is even less explored.~\footnote{
Monte Carlo simulations~\cite{Hanada:2009ne} 
may lead us to an understanding 
of M-theoretic aspects of the matrix theory.}
On the gauge theory side, 
for the ABJM theory, not even a non-perturbative formulation 
suitable for computer simulation is known, although
in the 't Hooft limit, 
a matrix model formulation~\cite{Hanada:2009hd} 
following the large-$N$ reduction technique introduced in~\cite{Ishii:2008ib} is available.  
However, experience with planar orbifold equivalences~\cite{Bershadsky:1998cb} 
suggests that it is not necessary to understand the full 
dynamics in order to prove the equivalence; in that case the proof was rather kinematical.  
Thus, orbifold equivalence may well be a good starting point to study the 
M-theoretic aspects of $AdS_4/CFT_3$ duality.
It seems to be possible to
relate our non-planar orbifold equivalence to 
a conjectured mirror symmetry in three dimensional 
gauge theories \cite{Intriligator:1996ex,Jensen:2009xh}:
the non-planar orbifold equivalence in this paper 
may be understood as a planar orbifold equivalence 
between the mirror theories~\cite{arXiv:1110.3803}. 
We hope to report progress in this direction in the near future.

\section*{Acknowledgement}
The authors would like to thank A.~Karch for stimulating discussions and comments. 
They also thank O.~Aharony, T.~Azeyanagi, S.~Hirano, S.~Kovacs, T.~Kuroki, 
S. Moriyama, N.~Obers, D.~Young 
and K. Zoubos for useful comments and discussion. 
The work of M.~H. is supported from Postdoctoral Fellowship for Research Abroad by Japan Society for the Promotion of Science.
C.~H. was supported in part by the U.S. Department of Energy under grant DE-FG02-96ER40956 and the Israel Science Foundation (grant number 1468/06).

\end{document}